\documentclass[aps,prb,amsmath,amssymb,reprint]{revtex4-2}
\usepackage{graphicx}
\usepackage{bm}

\begin{document}

\title{Destabilization of ferromagnetism by frustration
  and realization of a nonmagnetic Mott transition
  in the quarter-filled two-orbital Hubbard model}

\author{Katsunori Kubo}
\affiliation{Advanced Science Research Center,
  Japan Atomic Energy Agency, Tokai, Ibaraki 319-1195, Japan}

\date{\today}

\begin{abstract}
The two-orbital Hubbard model on a square lattice
at quarter filling (electron number per site $n=1$)
is investigated by the variational Monte Carlo method.
For the variational wave function,
we include short-range doublon-holon binding factors.
We find that the energy of this wave function is lower than
that of the density-density Jastrow wave function
partially including long-range correlations used in a previous study.
We introduce frustration to the model by
the next-nearest-neighbor hopping $t'$ in addition to
the nearest-neighbor hopping $t$.
For $t'=0$, a ferromagnetic state with staggered orbital order occurs
by increasing the Coulomb interaction $U$
before the Mott transition takes place.
By increasing $t'$, the region of this ferromagnetic phase shrinks,
and the Mott transition without magnetic order occurs.
\end{abstract}

\maketitle

\section{Introduction}
The Mott transition is one of the most remarkable phenomena
emerging from electron correlation.
According to band theory,
a system with an odd number of electrons per unit cell is a metal
since at least one band is partially filled
when we consider the spin degrees of freedom.
However, there are insulators
in which the electron number per unit cell is odd, e.g., MnO~\cite{Boer1937}.
To explain these insulators,
the importance of the Coulomb interaction between electrons
was suggested by Mott and Peierls~\cite{Mott1937,Mott1949} and
insulators in which the electron correlation plays a crucial role
for the realization of the insulating state are called Mott insulators.

To investigate the transition to the Mott insulator, the Mott transition,
the single-orbital Hubbard model at half filling,
i.e., electron number per site $n=1$, is a typical model.
When the onsite Coulomb interaction $U$ is much smaller than the bandwidth $W$,
the system should be in a metallic state.
On the other hand, for $U \gg W$,
most sites are occupied by a single electron
and the electrons rarely move due to the large Coulomb interaction.
Then, the system is expected to be insulating.
Thus, the Mott transition should take place at $U \simeq W$.
However, it is a hard task to describe the Mott transition theoretically
since it is a many-body problem.

To describe the electron correlation,
Gutzwiller proposed a variational wave function
[the Gutzwiller wave function (GWF)]~\cite{Gutzwiller1963}
and an approximation (the Gutzwiller approximation)~\cite{Gutzwiller1965}
to evaluate physical quantities in the GWF.
By applying the Gutzwiller approximation for $n=1$,
Brinkman and Rice pointed out that
the system becomes insulating at a finite value of $U$~\cite{Brinkman1970}.
The GWF for the single-orbital Hubbard model is given by
$\prod_{\bm{r}} [ 1-(1-g)n_{\bm{r}\uparrow}n_{\bm{r}\downarrow}]|\Phi \rangle$,
where $n_{\bm{r} \sigma}$ is the electron number operator of spin $\sigma$
at site $\bm{r}$ and $|\Phi \rangle$ is the ground state wave function
for $U=0$.
The Gutzwiller parameter $g$ tunes the probability
of double occupancy at each site.
If $g=0$, the double occupancy is completely prohibited
and electrons cannot move at all; that is, the system is insulating.
Brinkman and Rice found that $g$ becomes zero at a finite $U$
by employing the Gutzwiller approximation.

\begin{figure}
  \includegraphics[width=0.99\linewidth]
    {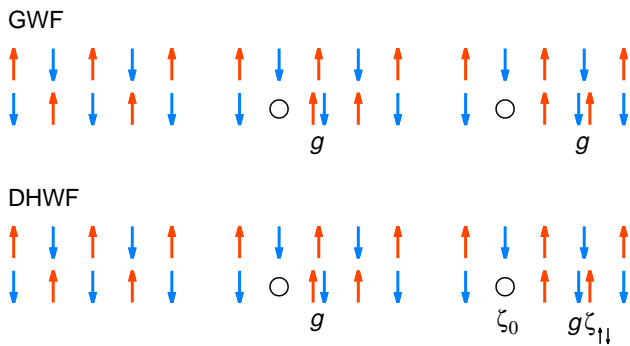}%
  \caption{
    Correlation factors for the GWF and DHWF.
    Here, we show them for the single-orbital model for simplicity.
    When a pair of a doublon and a holon is created,
    the factor $g$ appears (middle panels).
    In the GWF, the doublon and holon move freely without a change in the factor
    (top right panel).
    In the DHWF, the factors $\zeta_0$ and $\zeta_{\uparrow \downarrow}$
    appear when the pair separates (bottom right panel),
    where $0$ and $\uparrow \downarrow$ denote
    the holon and doublon states, respectively.
    In this figure, we have considered only the nearest-neighbor
    doublon-holon binding factors for clarity.
    \label{correlation_factors}}
\end{figure}

The complete suppression of double occupancy is an artifact
of the approximation.
As long as $U$ is finite, while the frequency may be low,
double occupancy should occur to reduce the kinetic energy.
To avoid introducing approximations, the variational Monte Carlo (VMC) method
was applied to the Hubbard model~\cite{Yokoyama1987JPSJ56_1490}.
Then, a finite $g$ was obtained even for a large $U$.
Moreover, the Mott transition disappeared in the VMC calculation;
that is, it was revealed that the GWF cannot describe the Mott transition
at least in one- and two-dimensional lattices.
The absence of the Mott transition in the GWF was also shown analytically
for the one-dimensional case~\cite{Metzner1988}.
The reason can be understood as follows.
In the GWF, when a pair of a doubly occupied site (doublon)
and an empty site (holon) is created,
the wave function is  multiplied by the factor $g$.
This factor does not change even when the distance between the doublon and holon
increases (top panels in Fig.~\ref{correlation_factors}).
In other words, the doublon and holon move freely after they are created.
For $U<\infty$, the numbers of doublons and holons are finite
and they can move freely; that is, the system is metallic, within the GWF.
Thus, the GWF cannot describe the Mott insulating state.

Later, it became clear that inclusion of intersite correlations
improves the situation~\cite{Yokoyama2002,Capello2006}.
Here, we consider the wave function including the doublon-holon binding factors
[doublon-holon binding wave function (DHWF)]~\cite{Kaplan1982,Yokoyama1990}.
In the DHWF, the wave function is multiplied by additional factors
when a doublon and a holon separate from nearest-neighboring sites
(bottom panels of Fig.~\ref{correlation_factors}).
It was shown that by an effect of these factors,
the DHWF can describe the Mott transition~\cite{Yokoyama2002}.

Another important issue of the Mott transition
is the competition with magnetism.
In the single-orbital Hubbard model for $n=1$
with only the nearest-neighbor hopping $t$,
the ground state is expected to be the antiferromagnetically ordered state
for $U>0$ due to the perfect nesting of the Fermi surface.
The unit cell contains two sites in the antiferromagnetic (AF) state; that is,
there are two electrons per unit cell.
Then, the AF state can be regarded as a band insulator.
To realize the Mott insulating state without ambiguity,
it is necessary to destabilize the antiferromagnetism.
A plausible way is to introduce frustration~\cite{Kashima2001,Morita2002,Parcollet2004,Watanabe2006,Mizusaki2006,Sahebsara2006,Kyung2006,Yokoyama2006,Onari2007,Clay2008}.
For example, it is found that the region of the AF phase shrinks
by increasing the next-nearest-neighbor hopping $t'$,
and the nonmagnetic Mott insulating phase appears~\cite{Kashima2001,Mizusaki2006,Yokoyama2006,Onari2007,Yamada2013}.

In this paper, based on the above developments in the single-orbital model,
we extend the research on the Mott transition to the two-orbital model.
We apply the VMC method to the two-orbital Hubbard model by employing the DHWF.
The interplay of the spin and orbital degrees of freedoms
leads to intriguing phenomena such as
orbital ordering and itinerant ferromagnetism,
and many theoretical studies have been conducted using this model.
To investigate the Mott transition, we consider integer values of $n$.
For $n=1$ without $t'$,
the ground state changes from the paramagnetic (PM) state
to a ferromagnetic (FM) state with staggered orbital order
by increasing the Coulomb interaction~\cite{Roth1966,Kugel1972,Kusakabe1993,
  Kusakabe1994,Momoi1998,Sakamoto2002,Kubo2009JPCS,Kubo2009PRB,Peters2010,
  Peters2011,Franco2018}.
It was found that partially spin polarized states are hard to
realize~\cite{Momoi1998,Sakamoto2002,Kubo2009JPCS,Kubo2009PRB,Franco2018},
and in the following, we consider only the completely spin polarized state,
in which only the majority spin states are occupied, as the FM state.
For $n=2$ without $t'$, the ground state is the AF state
without orbital order~\cite{Momoi1998,Kubo2009PRB,Franco2018}.
In both ordered phases, the electron number per unit cell is an even number,
and these phases can be regarded as band insulating phases.

The Mott transition has also been investigated for the two-orbital and
multiorbital Hubbard models~\cite{Bunemann1998,Han1998,Ono2003,Koga2006,
  Medici2011PRB,Medici2011PRL,Takenaka2012,Facio2017,Franco2018}
by assuming the PM phase.
In these studies,
the models with only the nearest-neighbor hopping are employed.
Then, if we allow for magnetic states,
the nonmagnetic Mott insulating phase disappears~\cite{Franco2018}.
Thus, we need to introduce frustration.

In the case of $n=2$,
the magnetic phase to be destabilized to realize the nonmagnetic Mott transition
is the AF phase,
and the situation may be similar to the single-orbital model.
In addition, the electron number per site is even,
and the distinction between the band insulator and Mott insulator
would be unclear.
Thus, in this study, we concentrate on the case of quarter filling $n=1$.
If an insulating state appears in the PM phase for $n=1$,
it is a Mott insulator without ambiguity.
We expect that by introducing frustration,
we can destabilize the FM phase
since this ferromagnetism is supported by the staggered ordering of
the orbital degrees of freedom.
Frustration usually destroys an AF state
and the destabilization of the ferromagnetism by frustration
is not so trivial and worth investigating.

\section{Model and wave functions}\label{model}
The two-orbital Hubbard model is given by
\begin{equation}
  \begin{split}
    H=&\sum_{\bm{k},\tau,\sigma}
    \epsilon_{\bm{k}}
    c^{\dagger}_{\bm{k} \tau \sigma}c_{\bm{k} \tau \sigma}
    +U \sum_{\bm{r}, \tau}
    n_{\bm{r} \tau \uparrow} n_{\bm{r} \tau \downarrow}\\
    &+U^{\prime} \sum_{\bm{r}}
    n_{\bm{r} 1} n_{\bm{r} 2}
    + J \sum_{\bm{r},\sigma,\sigma^{\prime}}
    c^{\dagger}_{\bm{r} 1 \sigma}
    c^{\dagger}_{\bm{r} 2 \sigma^{\prime}}
    c_{\bm{r} 1 \sigma^{\prime}}
    c_{\bm{r} 2 \sigma}\\
    &+J^{\prime}\sum_{\bm{r},\tau \ne \tau^{\prime}}
    c^{\dagger}_{\bm{r} \tau \uparrow}
    c^{\dagger}_{\bm{r} \tau \downarrow}
    c_{\bm{r} \tau^{\prime} \downarrow}
    c_{\bm{r} \tau^{\prime} \uparrow},
  \end{split}
\end{equation}
where 
$c^{\dagger}_{\bm{r} \tau \sigma}$ is the creation operator of the electron
with orbital $\tau$ ($=1$ or $2$)
and spin $\sigma$ ($=\uparrow$ or $\downarrow$)
at site $\bm{r}$,
$n_{\bm{r} \tau \sigma}=c^{\dagger}_{\bm{r} \tau \sigma}c_{\bm{r} \tau \sigma}$,
and
$n_{\bm{r} \tau}=\sum_{\sigma}n_{\bm{r} \tau \sigma}$.
$c^{\dagger}_{\bm{k} \tau \sigma}$ is the Fourier transform of
$c^{\dagger}_{\bm{r} \tau \sigma}$.
$U$ and $U'$ are intraorbital and interorbital Coulomb interactions,
respectively.
$J$ is the Hund's rule coupling and $J'$ denotes the pair-hopping interaction.
We use the relations $U=U'+J+J'$ and $J=J'$,
which hold in many orbitally degenerate systems~\cite{Tang1998}.

We consider the nearest-neighbor hopping $t$
and the next-nearest-neighbor hopping $t'$ on a square lattice.
Then, the kinetic energy is written as
\begin{equation}
  \epsilon_{\bm{k}}= -2 t (\cos k_x+ \cos k_y)-4 t' \cos k_x \cos k_y,
\end{equation}
where we have set the lattice constant as unity.
We can assume $t>0$ without loss of generality
since the sign of $t$ can be changed by the transformation
$c_{\bm{r} \tau \sigma} \rightarrow e^{i\bm{Q}\cdot\bm{r}}c_{\bm{r} \tau \sigma}$
[$\bm{Q}=(\pi,\pi)$]
without changing the other terms of the model.
Then, the bandwidth is $W=4t+4 \max(t,2|t'|)$.

The sign of $t'$ changes physical quantities in the PM phase
except for the electron-hole symmetric case $n=2$~\cite{Yokoyama2006}.
Preliminary calculations on an $8 \times 8$ lattice for $n=1$ indicate that
the FM state is stable against $t'<0$ in comparison with $t'>0$.
The completely spin-polarized FM state at $n=1$ has electron-hole symmetry,
and its energy does not depend on the sign of $t'$.
On the other hand, the energy of the PM state depends on the sign.
As inferred from the kinetic energy at $\bm{k}=(0,0)$,
$\epsilon_{\bm{k}=(0,0)}=-4t-4t'$,
the energy for $t'>0$ is lower than that for $t'<0$ at least
for a weak Coulomb interaction with $|t'| \ll t$ at low electron filling.
Thus, the PM solution may become advantageous for $t'>0$.
In the following, we consider only $t' \ge 0$ to destabilize the FM state.

Concerning the variational wave functions, we discuss three types:
the GWF, the DHWF, and
the wave function with the density-density Jastrow factor
used in a previous study~\cite{Franco2018}.

The GWF for the two-orbital model~\cite{Okabe1997,Bunemann1998,Kobayashi2006,Kubo2009JPCS,Kubo2009PRB,Kubo2011,Kubo2017} is given by
\begin{equation}
  | \Psi_{\text{GWF}} \rangle
  = P_{G} | \Phi \rangle,\\
\end{equation}
where $| \Phi \rangle$ is a one-electron wave function
which we will give below.
The Gutzwiller projection operator is defined as
\begin{equation}
  P_{G}
  =\prod_{\bm{r} \gamma}
  \left[ 1-(1-g_{\gamma})P_{\bm{r}\gamma} \right],
\end{equation}
where $\gamma$ denotes one of the 16 onsite states,
$P_{\bm{r} \gamma}$ is the projection operator
onto state $\gamma$ at site $\bm{r}$,
and $g_{\gamma}$ is a variational parameter.
In the following, $\gamma=0$ denotes the holon state, i.e., empty state.
Since the overall factor to the wave function is arbitrary,
we can omit one variational parameter.
In addition, by using the conservation of the number of electrons
for each spin and orbital and symmetry of the system,
we can reduce the number of $g_{\gamma}$ to be optimized to 5
in the PM state without ferro-orbital order.
Note that similar consideration reduces the number of Gutzwiller parameters
to one in the single-orbital Hubbard model.
There are four on-site states in the single-orbital model,
but one parameter can be omitted by considering
the overall factor to the wave function,
and two further parameters are omitted due to
the conservation of the up- and down-spin electrons.

The DHWF is given by
\begin{equation}
  | \Psi_{\text{DHWF}} \rangle
  = \prod_{i=1}^2 P^{(i)}_d P^{(i)}_h P_{G} | \Phi \rangle.
\end{equation}
$P^{(1)}_d$ is defined as
\begin{equation}
  P^{(1)}_d = \prod_{\bm{r}\, \gamma \in D}
  \left[1
    -(1-\zeta^{(1)}_{\gamma})P_{\bm{r} \gamma}
    \prod_{\bm{a}}(1-P_{\bm{r}+\bm{a} 0}) \right],
  \label{P_doublon}
\end{equation}
where $D$ denotes the set of doublon states,
i.e., onsite states with two electrons,
and $\bm{a}$ denotes the vectors connecting nearest-neighbor sites.
$P^{(1)}_d$ gives the factor $\zeta^{(1)}_{\gamma}$
when site $\bm{r}$ is in the doublon state $\gamma$
and there is no holon at nearest-neighbor sites $\bm{r}+\bm{a}$.
Similarly, $P^{(1)}_h$ is defined as
\begin{equation}
  P^{(1)}_h = \prod_{\bm{r}}
  \left[1
    -(1-\zeta^{(1)}_{0}) P_{\bm{r} 0}
    \prod_{\bm{a}\, \gamma\in D}(1-P_{\bm{r}+\bm{a} \gamma}) \right].
  \label{P_holon}
\end{equation}
The factor $\zeta^{(1)}_0$ appears
when a holon exists without a nearest-neighboring doublon.
By considering symmetry, four $\zeta^{(1)}_{\gamma}$ are independent
in the PM state without ferro-orbital order.
In this study, we also consider the next-nearest-neighbor hopping,
and we should include doublon-holon binding factors
for the next-nearest-neighbor sites.
They are represented by $P^{(2)}_d$ and $P^{(2)}_h$
and are defined similarly to $P^{(1)}_d$ and $P^{(1)}_h$, respectively,
by regarding $\bm{a}$ as the vectors connecting next-nearest-neighbor sites
and replacing $(1)$ with $(2)$ in Eqs.~\eqref{P_doublon} and \eqref{P_holon}.

We will compare some of our results with a previous study
using the density-density Jastrow wave function~\cite{Franco2018}:
\begin{equation}
  | \Psi_{\text{Jastrow}} \rangle
  = P_{J} | \Phi \rangle,
\end{equation}
with
\begin{equation}
  P_{J}=\exp
  \left( -\frac{1}{2}\sum_{\bm{r} \bm{r}' \tau \tau'} v^{\tau \tau'}_{\bm{r} \bm{r}'}
  n_{\bm{r} \tau}n_{\bm{r}' \tau'} \right).
\end{equation}
There are only two parameters for any distance,
$v^{11}_{\bm{r} \bm{r}'}=v^{22}_{\bm{r} \bm{r}'}$
and
$v^{12}_{\bm{r} \bm{r}'}=v^{21}_{\bm{r} \bm{r}'}$,
unless we consider ferro-orbital order.
However, this Jastrow wave function
partially includes the long-range correlations.
On the other hand,
the GWF and DHWF include only short-range correlations
but carefully treat them.

Without orbital order,
the one-electron part of the wave function $|\Phi\rangle$
is constructed by filling electrons inside the Fermi surface
defined by $\epsilon_{\bm{k}}$.
For an antiferro-orbital ordered state with ordering vector $\bm{Q}=(\pi,\pi)$,
we consider an effective Hamiltonian:
\begin{equation}
  H_{\bm{k} \tau}
  =
  \begin{pmatrix}
    \epsilon_{\bm{k}} & -\Delta_{\tau}\\
    -\Delta_{\tau} & \epsilon_{\bm{k}+\bm{Q}}\\
  \end{pmatrix},
\end{equation}
where
$\Delta_{\tau}
=
\Delta_{o}(\delta_{\tau 1}-\delta_{\tau 2})$.
We construct $|\Phi\rangle$ by filling electrons
from the bottom of the energy of this effective Hamiltonian.
$\Delta_o$ is a variational parameter.
$|\Phi\rangle$ is reduced to that without orbital order for $\Delta_o=0$.
As mentioned in the Introduction, at least for $t'=0$,
it is difficult to realize
a partially spin polarized FM state~\cite{Momoi1998,Sakamoto2002,Kubo2009JPCS,
  Kubo2009PRB,Franco2018}.
Thus, we consider only the majority-spin electrons in the FM state.

We optimize the variational parameters in each wave function
to reduce the expectation value of the energy
evaluated by the Monte Carlo method.
Physical quantities are also calculated by the Monte Carlo method
for the variational wave functions with the optimized variational parameters.

\section{Results}\label{results}
In the following,
we show the calculated results
for an $L \times L$ square lattice of $L=12$
with antiperiodic-periodic boundary conditions.
To examine the finite-size effect,
we also show some results for $L=10$ and $14$.
The number of electrons per site is fixed as $n=1$,
i.e., quarter-filling.
We have searched for the antiferro-orbital order in the PM phase
but we could not find it.
Thus, we show results only for the PM state without orbital order
and the FM state with antiferro-orbital order in the following.
Some results for $t'=0$ are compared with a previous study
using the density-density Jastrow wave function~\cite{Franco2018}
for the same lattice size with the same boundary conditions.

\begin{figure}
  \includegraphics[width=0.99\linewidth]
    {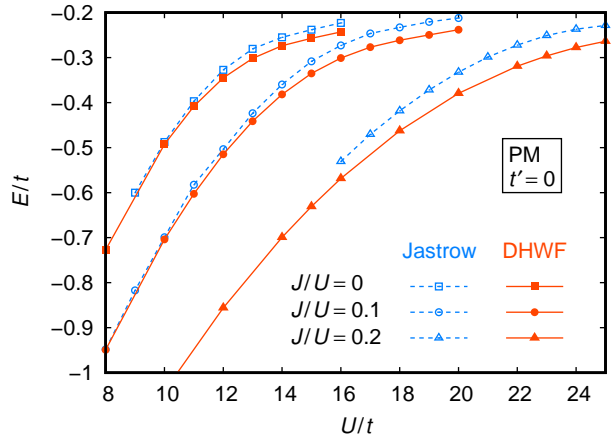}%
  \caption{
    Energy $E$ per site as a function of $U$
    for the Jastrow wave function (Ref.~\cite{Franco2018}, open symbols)
    and for the DHWF (solid symbols)
    for the Hund's rule coupling $J=0$ (squares), $J=0.1U$ (circles),
    and $J=0.2U$ (triangles).
    The next-nearest-neighbor hopping $t'$ is set to zero.
    \label{E_PM_t20}}
\end{figure}

First, we compare the energy of the PM state for $t'=0$
in the Jastrow wave function~\cite{Franco2018}
and the DHWF (Fig.~\ref{E_PM_t20}).
The energy of the DHWF is lower than
that of the Jastrow wave function in the entire range.
The lower energy indicates that the careful treatment
of the short-range correlations
is more important than the partial inclusion of the long-range correlations,
at least for $n=1$.
In addition, the DHWF includes the spin-dependent correlations;
for example,
doublons with parallel and antiparallel spin configurations
correspond to different variational parameters.
On the other hand, the density-density Jastrow wave function does not
include such spin dependence,
and the energy difference between these wave functions
becomes larger as the Hund's rule coupling $J$ increases.

\begin{figure}
  \includegraphics[width=0.99\linewidth]
    {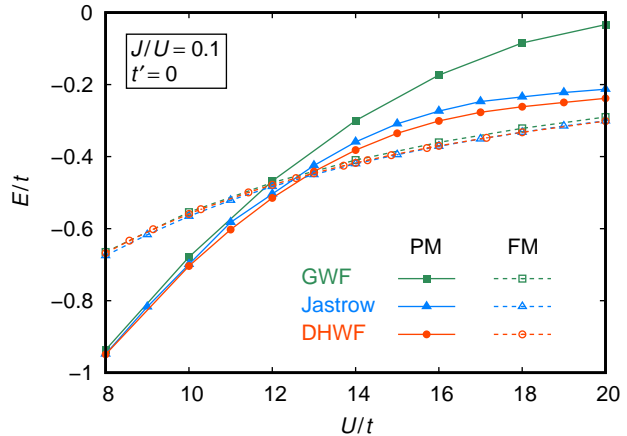}%
  \caption{
    Energy $E$ per site as a function of $U$
    in the PM (solid symbols) and FM (open symbols) states
    of the GWF (squares),
    of the Jastrow wave function (Ref.~\cite{Franco2018}, triangles),
    and of the DHWF (circles)
    for $J=0.1U$.
    The next-nearest-neighbor hopping $t'$ is set to zero.
    \label{E_PM_FM_t20_J.1U}}
\end{figure}

Figure~\ref{E_PM_FM_t20_J.1U} shows
the energy of the PM state without orbital order
and of the FM state with antiferro-orbital order
for the three kinds of wave functions for $J=0.1U$.
For the PM state, the GWF has much higher energy than
those of the Jastrow wave function and DHWF.
That is, the inclusion of the intersite correlations brings
a marked improvement in the PM state.
Between the wave functions with the intersite correlations,
as already mentioned above,
the DHWF has lower energy than the Jastrow wave function.
For the FM state,
the values of energy of these three wave functions are very close.
This closeness means that the intersite correlations are not very important
for the ordered phase in comparison with the PM phase.
For a large $U$, $U \gtrsim 16t$, the GWF has slightly
higher energy than the other wave functions.
For a smaller $U$, $U \lesssim 10t$, the Jastrow wave function has slightly
lower energy than the other wave functions.
For a smaller-$U$ case,
since electrons can move easily in comparison with a larger-$U$ case,
the long-range correlations play roles,
and the Jastrow wave function has an advantage.
We determine the FM transition point $U_{\text{FM}}$
by comparing the energy of the PM and FM states.
For $J=0.1U$ with $t'=0$, the FM transition occurs at $U_{\text{FM}} \simeq 13t$
in the DHWF.
In the Jastrow wave function, $U_{\text{FM}}/t = 12.5 \pm 0.5$
was reported~\cite{Franco2018}.
Since the values of energy for the DHWF and the Jastrow wave function are close
around the transition point,
these wave functions give close values of $U_{\text{FM}}$.

In the rest of this section, we show the results obtained with the DHWF.

\begin{figure}
  \includegraphics[width=0.99\linewidth]
    {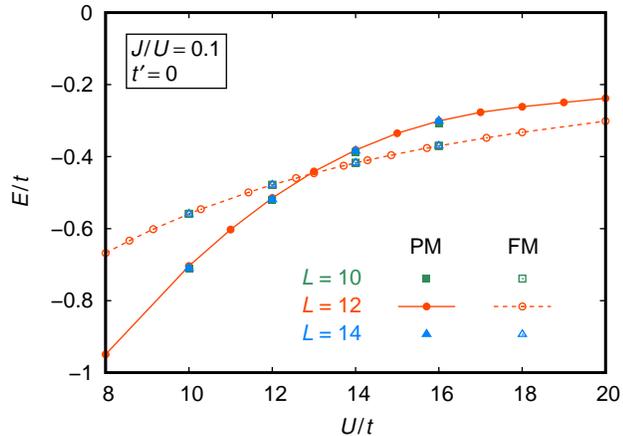}%
  \caption{
    Energy $E$ per site as a function of $U$
    in the PM (solid symbols) and FM (open symbols) states
    for $L=10$ (squares),
    for $L=12$ (circles),
    and for $L=14$ (triangles)
    for $J=0.1U$.
    The next-nearest-neighbor hopping $t'$ is set to zero.
    \label{E_PM_FM_t20_J.1U_size_dep}}
\end{figure}

To examine the finite-size effect,
we show the energy of PM and FM states for $L=10$, $12$, and $14$
as a function of $U$ for $J=0.1U$ in Fig.~\ref{E_PM_FM_t20_J.1U_size_dep}.
The finite-size effect on energy is very weak
and as a result,
the size dependence of $U_{\text{FM}}$ is weak.

\begin{figure}
  \includegraphics[width=0.99\linewidth]
    {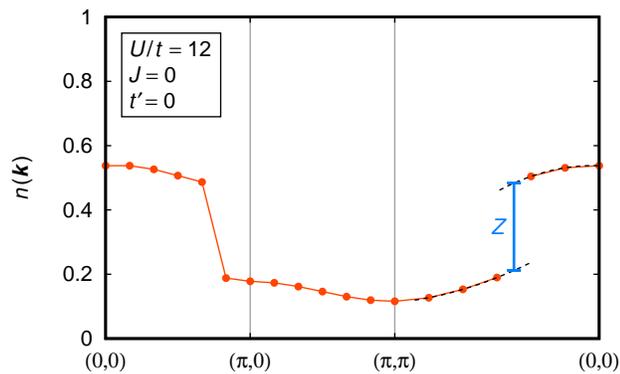}%
  \caption{
    Momentum distribution function
    $n(\bm{k})=
    \langle c^{\dagger}_{\bm{k} \tau \sigma} c_{\bm{k} \tau \sigma} \rangle$
    for $U=12t$, $J=0$, and $t'=0$ in the PM phase.
    The renormalization factor $Z$ is estimated
    by extrapolating $n(\bm{k})$
    from above and below the Fermi momentum
    along $(\pi,\pi)$--$(0,0)$ (dashed lines).
    Due to the antiperiodic boundary condition for the $x$ direction,
    we shift $k_x$ by $\pi/L$ ($L=12$); for example,
    $(\pi,\pi)$ denoted in this figure actually means
    the point $(\pi-\pi/L,\pi)$.
    \label{nk_PM_t20_U12_J0}}
\end{figure}

Figure~\ref{nk_PM_t20_U12_J0} shows
the momentum distribution function
$n(\bm{k})= \langle c^{\dagger}_{\bm{k} \tau \sigma} c_{\bm{k} \tau \sigma} \rangle$
for $U=12t$, $J=0$, and $t'=0$ in the PM phase,
where $\langle \cdots \rangle$ denotes the expectation value
in the optimized wave function.
This quantity does not depend on the orbital and spin in the PM phase.
Due to the correlation effect,
the jump at the Fermi momentum is reduced from unity.
We define the renormalization factor $Z$
by this jump along $(\pi,\pi)$--$(0,0)$.
$Z$ is inversely proportional to the effective mass,
and in an insulating state $Z=0$.
To estimate $Z$ in a finite-size lattice,
we extrapolate $n(\bm{k})$ from above and below the Fermi momentum,
as shown in Fig.~\ref{nk_PM_t20_U12_J0}.

\begin{figure}
  \includegraphics[width=0.99\linewidth]
    {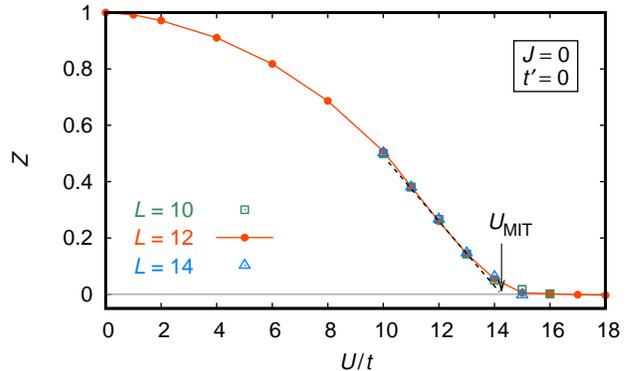}%
  \caption{
    Renormalization factor $Z$ as a function of $U$
    for $J=0$ and $t'=0$
    in the PM phase
    for $L=10$ (squares),
    for $L=12$ (circles),
    and for $L=14$ (triangles).
    Here, we estimate $U_{\text{MIT}}$ by linearly extrapolating $Z$ to zero
    from data of $Z>0.1$ for $L=12$.
  \label{Z_PM_t20_J0_size_dep}}
\end{figure}

In Fig.~\ref{Z_PM_t20_J0_size_dep},
we show the $U$ dependence of $Z$ for $t'=0$ and $J=0$ in the PM phase.
$Z$ is reduced from unity by increasing $U$
and seems to vanish at $U \simeq 14t$.
From the evaluation procedure of $Z$,
it is difficult to determine $Z$ accurately for a finite-size lattice
when $Z$ is small.
The number of data points available to determine $Z$ is $O(L)$
for an $L \times L$ lattice,
and it may be difficult to determine $Z$ smaller than $1/L$.
Indeed, we observe the size dependence of $Z$ for $U \simeq 14t$.
Thus, here, we determine the metal-insulator transition point $U_{\text{MIT}}$
by linearly extrapolating data with $Z>0.1$ to $Z=0$
as shown in Fig.~\ref{Z_PM_t20_J0_size_dep}.
By using different initial variational parameters,
we estimate errors in $Z$ as $\delta Z \lesssim 0.01$ for $Z \simeq 0.1$
in the present calculation.
Then, the error in $U_{\text{MIT}}$ due to the extrapolation is estimated
to be $\delta U_{\text{MIT}} \lesssim 0.1t$.
Note that the error from the finite-size effect would be larger.
To evaluate the value of $U_{\text{MIT}}$ more accurately,
we should carefully check the lattice-size dependence by using larger lattices.
It is outside the scope of this study.
In the density-density Jastrow wave function, the transition point
was estimated as $U_{\text{MIT}}/t = 13 \pm 1$~\cite{Franco2018}.

By determining $U_{\text{FM}}$ and $U_{\text{MIT}}$ for each $t'$,
we construct phase diagrams.
\begin{figure}
  \includegraphics[width=0.99\linewidth]
    {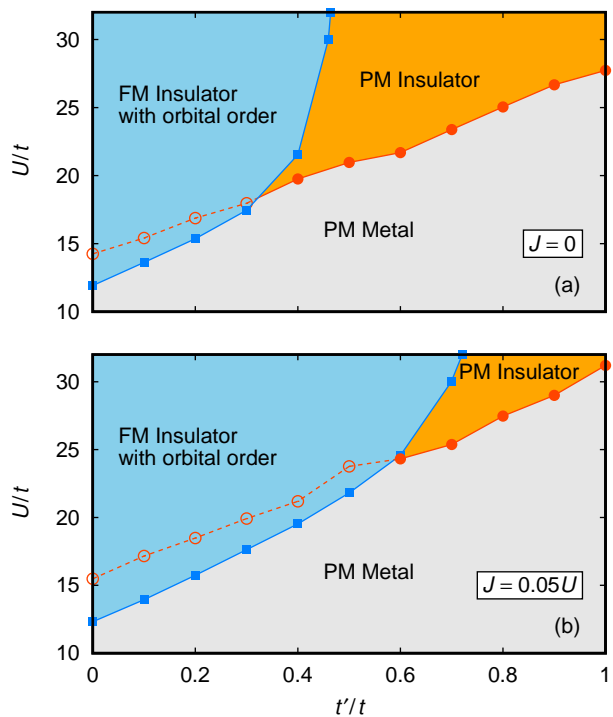}%
  \caption{
    Phase diagrams for the two-orbital Hubbard model
    with the next-nearest-neighbor hopping $t'$
    (a) for $J=0$ and (b) for $J=0.05U$.
    The open circles with the dashed lines indicate
    the metal-insulator transition in the PM phase
    when we ignore the FM state.
  \label{phase_diagram_J0_J.05U}}
\end{figure}
For $J=0$ [Fig.~\ref{phase_diagram_J0_J.05U}(a)],
we obtain $U_{\text{MIT}} \simeq 14t$ for $t'=0$
in the PM phase (see Fig.~\ref{Z_PM_t20_J0_size_dep}),
but the PM-FM transition already occurs at $U_{\text{FM}}\simeq 12t$.
Thus, the nonmagnetic Mott transition does not occur for $t'=0$.
By increasing frustration $t'$,
the staggered order of the orbital would be destabilized.
Indeed, we find that the FM insulating phase with orbital order shrinks,
and the nonmagnetic Mott transition emerges at $t' \gtrsim 0.3t$.
For $J=0.05U$ [Fig.~\ref{phase_diagram_J0_J.05U}(b)],
the phase diagram is similar to that for $J=0$ at $t' \lesssim 0.3t$.
By an effect of the Hund's rule coupling,
which is expected to stabilize magnetic phases,
the region of the FM phase is extended, in particular, for $t' \gtrsim 0.3t$.

For $J=0$, antiferro-orbital order without ferromagnetism may be expected,
but we could not find such a phase.
Instead, we found the FM state with antiferro-orbital order even for $J=0$.
In this state, each orbital state is ferromagnetic
due to the strong correlation effect within each orbital.
Note that this result does not mean the realization of
a FM state in the single-orbital Hubbard model for $n=0.5$.
The realization of the ferromagnetism in the two-orbital Hubbard model
is supported by the antiferro-orbital order.
For $J=0$, this FM state
is equivalent to an antiferro-orbital ordered state
in which the $\tau=1$ orbital states are occupied with only up-spin electrons
and the $\tau=2$ orbital states are occupied with only down-spin electrons.
For a finite $J$, this state should have higher energy than the FM state.

\section{Summary}\label{summary}
We have investigated the quarter-filled two-orbital Hubbard model
on a square lattice with next-nearest-neighbor hopping $t'$
using the VMC method.

In the variational wave function DHWF,
we have considered the nearest-neighbor and next-nearest-neighbor
doublon-holon binding factors.
We have found that the energy for the PM state of the DHWF
is lower than that of the density-density Jastrow wave function
used in a previous study~\cite{Franco2018}.
This result means that a careful treatment of the short-range correlations is
more important than partial inclusion of the long-range correlations,
at least for $n=1$.

In the ordinary two-orbital Hubbard model
with the only nearest-neighbor hopping $t$,
the FM transition with staggered orbital order
occurs before the Mott transition when we increase the Coulomb interaction $U$.
To realize the nonmagnetic Mott transition,
it is necessary to suppress the FM phase.
For this purpose,
we have introduced frustration to the model
by the next-nearest-neighbor hopping $t'$,
which is expected to destabilize the staggered orbital order
supporting the FM state.
We have found that the region of the FM phase is, indeed, shrunk by $t'$
and the nonmagnetic Mott transition occurs.
Thus, it is revealed that the realization of the nonmagnetic Mott transition
is not limited to the simple single-orbital case and
the research field of Mott physics can be extended to multiorbital systems.

We also searched for the antiferro-orbital order in the PM phase,
but we could not find it,
while it was reported
when using a dynamical mean-field theory~\cite{Peters2011}.
It is interesting to realize such a pure orbital order
by improving the model and/or the wave function
for a finite-dimensional lattice.
In particular, if such a phase were to appear between
the PM insulator (Mott insulator) and the FM insulator with orbital order,
the microscopic description of successive transitions between these phases
would be intriguing.
This phase is an important future problem.


%

\end{document}